\documentclass[PRD,twocolumn,showpacs,superscriptaddress,preprintnumbers,nofootinbib,amsmath,amssymb]{revtex4}

\usepackage{graphicx}
\usepackage{dcolumn}
\usepackage{bm}

\def\vecp{\bm{p}}
\def\vecv{\bm{v}}
\def\vecx{\bm{x}}

\newcommand{\beq}{\begin{equation}}
\newcommand{\eeq}{\end{equation}}
\newcommand{\beqa}{\begin{eqnarray}}
\newcommand{\eeqa}{\end{eqnarray}}

\newcommand{\thetav}{{\theta_v}}
\newcommand{\meff}{{m_{\mathrm{eff}}}}
\newcommand{\gfactor}{{(1-\bm v \cdot \hat{\bm p})}}

\begin{document}

\title{Neutrino Oscillations, Lorentz/\textit{CPT} Violation, and Dark Energy}

\author{Shin'ichiro Ando}
\affiliation{California Institute of Technology, Mail Code 350-17,
     Pasadena, California 91125, USA}
\author{Marc Kamionkowski}
\affiliation{California Institute of Technology, Mail Code 350-17,
     Pasadena, California 91125, USA}
\author{Irina Mocioiu}
\affiliation{Pennsylvania State University, 104 Davey Lab,
     University Park, Pennsylvania 16802, USA}

\date{October 22, 2009}

\begin{abstract}
If dark energy (DE) couples to neutrinos, then there may
be apparent violations of Lorentz/{\it CPT} invariance in neutrino
oscillations.  The DE-induced Lorentz/{\it CPT} violation takes
a specific form that introduces neutrino oscillations that
are energy independent, differ for particles and antiparticles,
and can lead to novel effects for neutrinos propagating through
matter.  We show that ultra-high-energy neutrinos may provide
one avenue to seek this type of Lorentz/{\it CPT} violation in
$\nu_\mu$-$\nu_\tau$ oscillations, improving the current
sensitivity to such effects by seven orders of magnitude.
Lorentz/{\it CPT} violation in
electron-neutrino oscillations may be probed with the
zenith-angle dependence for high-energy atmospheric neutrinos.
The ``smoking gun,'' for DE-neutrino coupling would,
however, be a dependence of neutrino oscillations on the direction of the
neutrino momentum relative to our peculiar velocity with respect
to the CMB rest frame.  While the amplitude of this directional
dependence is expected to be small, it may nevertheless be worth
seeking in current data and may be a target for future neutrino
experiments.
\end{abstract}


\maketitle

\section{Introduction}

The accelerated cosmic expansion \cite{SNIa} poses difficult
questions for theoretical
physics~\cite{Copeland:2006wr,Caldwell:2009ix,Silvestri:2009hh}.
Is it simply due to a cosmological constant?  Is some new
negative-pressure dark energy (DE) required?  Is general
relativity modified at large distance scales?  
The major thrust of the empirical assault on these questions has been
to determine whether the expansion history and growth of
large-scale structure are consistent with a cosmological
constant or require something more exotic \cite{DETF}.  

However, it may be profitable to explore whether there are
other experimental consequences of the new physics---which we
collectively refer to as DE, although it may
involve a modification of gravity rather than the introduction
of some new substance---responsible for accelerated expansion.
If cosmic acceleration is due to a cosmological constant (i.e.,
if general relativity is valid and the
equation-of-state parameter is $w=-1$), then 
the vacuum is Lorentz invariant.  If, however, something else is
going on, then the ``vacuum'' has a preferred
frame: the rest frame of the cosmic microwave background (CMB).  If,
moreover, dark energy couples somehow to standard-model
particles, then there may be testable (apparent) violations of
Lorentz invariance.  For example, if DE is coupled to the
pseudoscalar $F\tilde F$ of electromagnetism
\cite{Carroll:1998zi}, there may be a ``cosmological
birefringence'' that rotates the linear polarization of
cosmological photons; CMB searches for such a rotation
\cite{Lue:1998mq} constrain this rotation to be less than a few
degrees~\cite{Feng:2006dp}.

Here we explore DE-induced Lorentz/{\it CPT}-violating
effects in the neutrino sector.  We show that the form of a
Lorentz-violating coupling between neutrinos and dark energy is
highly restricted under fairly general assumptions.\footnote{The
coupling of neutrinos to dark energy has also been considered in
the context of ``mass-varying neutrinos''~\cite{MaVaN}, but that
implementation of the DE-neutrino coupling does not lead to the
type of Lorentz/{\it CPT}-violating effects we discuss here.}
The coupling engenders an additional source for neutrino mixing
(e.g., Ref.~\cite{Gu:2005eq}), resulting in neutrino
oscillations with a different energy dependence than vacuum
oscillations and different oscillation probabilities for neutrinos
and antineutrinos.  While similar Lorentz/{\it CPT}-violating
oscillations have been considered
before~\cite{bargeretal,CPT_nu,Kostelecky2004}, we emphasize
here that cosmic acceleration dictates a specific form for such
effects.

Data from Super-Kamiokande and K2K~\cite{Gonzalez-Garcia2004} and
AMANDA/IceCube~\cite{Abbasi2009} already tightly constrain {\it
CPT}-violating parameters for $\nu_\mu$-$\nu_\tau$ mixing, and
those from solar-neutrino experiments and
KamLAND~\cite{Bahcall2002} do so for $\nu_e$-$\nu_\mu$ mixing.
However, the effects of DE-induced {\it CPT} violation
become more significant at
higher energies \cite{Hooper:2005}.  Here we show that
next-generation measurements of ultra-high-energy neutrinos
produced by spallation of ultra-high-energy cosmic rays
will increase the sensitivity to {\it CPT}-violating
$\nu_\mu$-$\nu_\tau$ oscillations by seven orders of magnitude.
We also show that these {\it CPT}-violating
couplings may lead to novel effects in the zenith-angle
dependence for atmospheric neutrinos in the $\sim100$~GeV range.

While such {\it CPT}-violating effects, if detected, could be
attributed simply to intrinsic {\it CPT} violation in
fundamental physics, not related to DE, a
DE-neutrino coupling further predicts a directional effect: the
neutrino-mixing parameters depend on the neutrino
propagation direction relative to our peculiar velocity with
respect to the CMB rest frame.  While this signature will likely
remain elusive even to next-generation experiments, it would, if
detected, be a ``smoking gun'' for DE beyond a
cosmological constant.  It is therefore worth considering as
a long-range target for future neutrino experiments.  It may
also be worthwhile to search current data in case an implementation
of DE-neutrino coupling different from that we consider here
leads to a different energy dependence for these directional
effects.  We therefore work out explicitly the directional
dependence to aid experimentalists who may wish to look for such
correlations in current data.

Below, we first derive in Sec.~\ref{sec:formalism} the form of
the Lorentz/{\it CPT} violation
allowed by a DE-neutrino coupling and discuss the resulting
neutrino-oscillation physics.
In Sec.~\ref{sec:cosmogenic} we apply the formalism to
cosmogenic ultra-high-energy neutrinos, and obtain projected 
sensitivities of future detectors to these effects in
$\nu_\mu$-$\nu_\tau$ oscillations.  In Sec.~\ref{resonance} we
discuss matter-induced effects for $\nu_e$
oscillations in high-energy atmospheric neutrinos in the
presence of Lorentz-invariance--violating mixings.
Concrete formulas for the directional dependence on oscillation
probabilities are given in Sec.~\ref{sec:direction}.
Finally, we discuss some theoretical implications in 
Sec.~\ref{implications} and summarize and conclude in
Sec.~\ref{sec:conclusion}.

\section{The dark-energy--neutrino coupling}
\label{sec:formalism}

\subsection{General Formalism}

Following Ref.~\cite{Kostelecky2004}, 
the neutrino fields are denoted by Dirac spinors $\{\nu_e,
\nu_\mu,\nu_\tau,\cdots\}$ and their charge conjugates by
$\{\nu_{e^c}, \nu_{\mu^c}, \nu_{\tau^c},\cdots\}$, where
$\nu_{x^c}\equiv \nu_x^c \equiv C \bar \nu_x^T$ is the charge-conjugated
spinor, and $C$ is the charge-conjugation matrix.  The $2N$ 
fields (where $N$ is the number of flavors) and their conjugates are
arranged in a single object $\nu_A$ where $A$ ranges over
$e,\mu,\tau,\cdots,e^c,\mu^c,\tau^c,\cdots$.

With a canonical kinetic term in the neutrino Lagrangian,
the most general Lorentz/{\it CPT}-violating Dirac equation
is,\footnote{Additional possibilities arise with a non-canonical
kinetic term; we comment briefly on possible consequences below.}
\begin{equation}
     (i\gamma^\mu \partial_\mu -M_{AB}) \nu_B=0,
\end{equation}
where
\begin{eqnarray}
     M_{AB} &\equiv& m_{AB} + i m_{5AB}\gamma_5 + a^\mu_{AB}
     \gamma_\mu + b^\mu_{AB} \gamma_5 \gamma_\mu
     \nonumber\\&&{}
     + \frac{1}{2} H^{\mu\nu}_{AB} \sigma_{\mu\nu}.
\label{eqn:mostgeneral}
\end{eqnarray}
The usual mass terms are $m+im_5\gamma_5 \equiv m_L P_L + m_R
P_R$, where $m_R=(m_L)^\dagger = m+im_5$, $P_L=(1-\gamma_5)/2$,
and $P_R=(1+\gamma_5)/2$.  The $2N\times 2N$ mass matrix $m_R$
is written in terms of $N\times N$ matrices $L$, $R$, and $D$,
through
\begin{equation}
     m_R = \left( \begin{array}{cc} L & D \\ D^T &
     R \end{array} \right).
\end{equation}
Here, $R$ and $L$ are the right- and left-handed Majorana
neutrino masses ($L=0$ is required if electroweak gauge
invariance is preserved), and $D$ is the Dirac-mass matrix.  The
$R$ and $L$ matrices are required to be symmetric, and $R$, $L$,
and $D$ can most generally be complex.

\subsection{Dark-energy--induced Lorentz violation}

Lorentz violation in Eq.~(\ref{eqn:mostgeneral}) is parametrized
by the four-vectors $a^\mu$, $b^\mu$, and the antisymmetric
tensor $H^{\mu\nu}$.  The parameters $a^\mu$ and $b^\mu$ are
both {\it CPT} and Lorentz violating, while $H^{\mu\nu}$ is
Lorentz violating but {\it CPT} conserving.  
While these parameters are non-zero for the most general
Lorentz/{\it CPT}-violating Dirac equation
\cite{Kostelecky2004}, the allowable forms for $a^\mu$, $b^\mu$,
and $H^{\mu\nu}$ are highly restricted if the Lorentz/{\it CPT}
violation is induced by coupling to dark energy.

The smallness of the CMB quadrupole demands that the
3-dimensional hypersurfaces of constant DE density must be
closely aligned with those of constant CMB temperature
\cite{Erickcek:2008jp}.  The preferred frame associated with the
cosmic expansion is then parametrized by a unit four-vector
$l^\mu$ which is orthogonal to surfaces of constant CMB
temperature; i.e., in the CMB rest frame, it is
$l^\mu=(1,0,0,0)$.  The symmetry of the problem thus dictates
that $a^\mu\propto l^\mu$ and $b^\mu\propto l^\mu$.  The tensor
$H^{\mu\nu}$ is antisymmetric, and there is no way to 
construct an antisymmetric tensor $H^{\mu\nu}$ from a single
four-vector; we thus expect $H^{\mu\nu}=0$ for DE-neutrino
coupling.

Furthermore, since neutrinos are produced and interact in weak
eigenstates, it is only the combination $(a_L)^\mu_{ab} \equiv
(a+b)^\mu_{ab}$ that is relevant for neutrino phenomenology.
{\it Thus, the Lorentz/{\it CPT} violation induced in neutrino
physics can be parametrized entirely by a single
four-vector-valued $(a_L)^{\mu}_{ab} \propto l^\mu$ matrix in
the flavor space.}  

\subsection{Neutrino Oscillations}
\label{sec:oscillation}

The propagation of the flavor eigenstates is then described by
an effective Hamiltonian
\begin{widetext}
\begin{equation}
     (h_{\mathrm{eff}})_{ab} = \left( \begin{array}{cc} p\delta_{ab} +
     (\tilde m^2)_{ab}/2p + (a_L)^\mu_{ab} p_\mu/p & 0 \\
     0 & p\delta_{ab} + (\tilde m^2)^*_{ab}/2p - (a_L)^{\ast\mu}_{ab}
     p_\mu/p \end{array} \right),
\label{eqn:hamiltonian}
\end{equation}
\end{widetext}
where the flavor indices $a$ and $b$ run over the flavor
eigenstates $e$, $\mu$, and $\tau$ and $e^c$, $\mu^c$, and $\tau^c$.
Here, $p \equiv |\vecp|$, with $\vecp$ the neutrino momentum,
and $\tilde m^2\equiv m_l m_l^\dagger$ is the usual mass matrix,
with $m_l=L - DR^{-1} D^T$.

Equation~(\ref{eqn:hamiltonian}) has several implications:  (i)
Since the matrix is block-diagonal, there is no mixing between
neutrinos and antineutrinos (as may arise in
more general Lorentz-violating scenarios~\cite{Kostelecky2004}).
(ii) Since $a_L$ appears with
opposite sign in the neutrino and antineutrino entries in the
Hamiltonian, a nonzero $a_L$ implies (apparent) {\it CPT}
violation---i.e., the propagation of neutrinos and antineutrinos
is not the same.  Thus, for example, if the anomalous LSND
results had stood, the {\it CPT}-violating explanations (e.g.,
Ref.~\cite{deGouvea:2006qd}) for them \cite{LSND} 
may have implied DE-neutrino coupling.  (iii) The
mixing induced by DE-neutrino coupling is energy independent
(like in the Mikheev-Smirnov-Wolfenstein, or MSW, effect~\cite{MSW}), as
opposed to vacuum mixing, which declines as $E^{-1}$.
Thus, these effects will become increasingly visible at higher
energies.  The detailed form of {\it CPT} violation implied by
this effect is also thus different than that obtained with
different $\Delta m^2$ for neutrinos and antineutrinos.  (iv)
There may also be novel effects for neutrinos propagating
through matter, an effect we discuss further in
Sec.~\ref{resonance} below.

Finally, (v) the neutrino oscillations induced by DE-neutrino coupling
are {\it frame dependent}.  If the observer is in the rest frame
of the CMB, then $(a_L)^\mu p_\mu \propto E$, and neutrino
oscillations are independent of the neutrino direction.
However, the Solar System moves with respect to the CMB rest
frame with a velocity $v \simeq 370$~km~s$^{-1}$.
DE-induced neutrino oscillations will therefore depend on
$(a_L)^\mu p_\mu \propto E (1-\vecv \cdot \hat{\vecp})$, where
$\hat{\vecp}$ is the neutrino-propagation direction and $\vecv$
is our peculiar velocity with respect to the CMB rest frame.
There will thus be an annual modulation in solar-neutrino
oscillations, a diurnal modulation in laboratory neutrino-mixing
experiments, and a direction dependence in oscillations of
cosmogenic neutrinos.

Since neutrino mixing arises only as a consequence of the {\it
traceless} part of the propagation Hamiltonian, the DE-neutrino
coupling must (like the vacuum mass matrix) be flavor-violating if 
neutrino oscillations are to be affected.

\subsection{Two-flavor oscillations}
\label{sec:two-flavor}

The evolution equation for DE-induced two-flavor mixing is of
the form,
\begin{widetext}
\begin{equation}
   i \frac{d}{dt} \left( \begin{array}{c} \nu_a \\
   \nu_b \end{array} \right) = \frac{1}{2}
   \left( \begin{array}{cc} -\frac{\Delta m^2}{2E} \cos 2\thetav
   - \meff\gfactor \cos 2\theta_d & \frac{\Delta m^2}{2E}
   \sin2\thetav +\meff\gfactor \sin2\theta_d e^{i\eta}\\
   \frac{\Delta m^2}{2E} \sin2\thetav +\meff\gfactor
   \sin2\theta_d e^{-i\eta} & \frac{\Delta m^2}{2E} \cos 2\thetav +
   \meff\gfactor\cos2\theta_d \end{array} \right)
   \left( \begin{array}{c} \nu_a \\ \nu_b \end{array} \right),
\label{eqn:DEnumixinggeneral}
\end{equation}
\end{widetext}
where $\meff$ is an effective mass parameter, and $\theta_d$ and $\eta$
are a mixing angle and relative phase in the DE-neutrino coupling
matrix, respectively.  There is also the usual vacuum mass
difference (squared) $\Delta m^2$ and the vacuum mixing angle
$\theta_v$.  The analogous propagation equations for
antineutrinos are the  same as Eq.~(\ref{eqn:DEnumixinggeneral})
with the replacements $\meff \rightarrow - \meff$ and $\eta \to
-\eta$, the changes in sign a manifestation of {\it CPT} violation.

Recall that if the propagation Hamiltonian is of the form,
\begin{equation}
     h = M \left( \begin{array}{cc} -\cos2\theta & \sin
     2\theta \\ \sin 2\theta & \cos 2\theta \end{array}
     \right),
\label{eqn:simplematrix}
\end{equation}
then the probability for one species of neutrino to convert to a
different neutrino after a distance $L$ is
\begin{equation}
     P(\nu_a \rightarrow \nu_b) = \sin^2 2\theta \sin^2(ML).
\end{equation}
Here we have neglected the {\it CP}-violating phase in
Eq.~(\ref{eqn:simplematrix}) because it does not affect the oscillation
probability.
The propagation Hamiltonian in
Eq.~(\ref{eqn:DEnumixinggeneral}) can be written in the form of
Eq.~(\ref{eqn:simplematrix}) with the following
relations~\cite{Coleman1999}:
\begin{widetext}
\begin{eqnarray}
     M^2 &=& \left(\frac{\Delta m^2}{4E}\right)^2 +
      \frac{\meff^2\gfactor^2}{4} 
      + \frac{\Delta m^2}{4E}\meff \gfactor 
      (\cos 2\theta_v \cos 2\theta_d + \sin 2\theta_v \sin 2\theta_d
      \cos\eta),
      \\
     \sin^2 2\theta &=& \frac{1}{M^2}\left[ \left(\frac{\Delta
	 m^2}{4E}\right)^2\sin^22\thetav
     +
     \frac{\meff^2\gfactor^2}{4}\sin^22\theta_d + \frac{\Delta
     m^2}{4E}\meff\gfactor\sin 2\thetav \sin2\theta_d \cos\eta\right] .
\end{eqnarray}
\end{widetext}
Note that this time $\sin 2\theta$ does indeed depend on $\eta$, as it is not
the overall phase, but the {\it relative} one, that cannot be rotated
away by redefinition of wave functions.
The oscillation length is then $L_{\mathrm{osc}}=\pi M^{-1}$.
In the absence of DE-neutrino coupling, we recover the
standard oscillation length $L_{\rm osc}=4 \pi E/\Delta m^2$ and mixing
angle $\theta=\thetav$.  If $\meff \gg \Delta m^2/2E$, then the
oscillation length is $L_{\rm osc}=2\pi\meff^{-1}\gfactor^{-1}$.

In general, $m_{\rm eff}$ can be either positive or negative.
However, from the symmetry of the Hamiltonian, the relevant
parameter space can be limited to $m_{\rm eff} \ge 0$, $0 \le \theta_{d}
\le \pi/4$, and $0 \le \eta \le \pi$~\cite{Gonzalez-Garcia2004}.

Thus far, no deviations from standard three-flavor neutrino
oscillations have been 
discovered in experimental data (except LSND~\cite{LSND}), and this
yields constraints on {\it CPT}-violating parameters, especially for
$m_{\rm eff}$.  By analyzing solar-neutrino and KamLAND data,
Ref.~\cite{Bahcall2002} obtained an upper limit of $m_{\rm eff}
< 3.1 \times 10^{-20}$ GeV for $\nu_e$-$\nu_\mu$ mixing.
Atmospheric and accelerator data provide an upper limit for
$\nu_\mu$-$\nu_\tau$ mixing of $m_{\rm 
eff} < 5 \times 10^{-23}$~GeV~\cite{Gonzalez-Garcia2004}.

\section{Ultra-high-energy neutrinos}
\label{sec:cosmogenic}

\subsection{Prediction}

Given that DE-induced neutrino mixing becomes increasingly
important, relative to vacuum mixing, at high energies, 
the DE-neutrino coupling can be probed with ultra-high-energy
cosmogenic neutrinos.  These neutrinos are produced by the
interaction of ultra-high-energy cosmic-ray protons with CMB
photons~\cite{BZ}:
\begin{equation}
     p\gamma \to n \pi^+ \to n \mu^+ \nu_\mu \to n e^+ \nu_e \nu_\mu
     \bar\nu_\mu .
\end{equation}
The fact that the Greisen-Zatsepin-Kuzmin cutoff~\cite{GZK} has
now been observed by the HiRes~\cite{HiRes} and
Auger~\cite{Auger_GZK} collaborations implies that this
interaction must be occurring.  And if so, there must be a
population of cosmogenic neutrinos with energies
$10^{17}$--$10^{20}$ eV~\cite{BZ,GZK_nu,WB,Yuksel2007}.

The characteristic distance between the source of these
neutrinos and the Earth is the Hubble distance $cH_0^{-1}$,
which is much longer than the oscillation length---i.e.,
$cH_0^{-1} \gg M^{-1}$---as long as $m_{\rm eff} \gg H_0 =
10^{-42}$ GeV, as is always the case here.
Therefore, any oscillatory features in neutrino mixing will be
washed out; the probability for conversion of a
cosmogenic neutrino from its production flavor to another flavor
en route from the source is then simply $\sin^2 2\theta / 2$.
Cosmogenic neutrinos mostly originate from pion decays, with the
characteristic flavor ratio $\nu_e$:$\nu_\mu$:$\nu_\tau =
1$:2:0.
The result of standard vacuum mixing would be a flavor ratio at the Earth of $\nu_e$:$\nu_\mu$:$\nu_\tau = 1$:1:1. 
While possible corrections to this flavor ratio can be induced by small three-flavour oscillation effects or other new physics, here we concentrate on exploring the consequences of the DE-induced mixing.

For the sake of simplicity, we focus on
$\nu_\mu$-$\nu_\tau$ mixing (and their antiparticles).
In the absence of a DE-neutrino interaction, these two flavors
are maximally mixed; i.e., $\theta_v = \pi/4$, and thus even if only
$\nu_\mu$ are produced at the source, an equal number of $\nu_\mu$
and $\nu_\tau$ is generated by mixing.  However, this can be
altered if there is a DE-neutrino interaction.
The flux of $\nu_\mu$ and $\nu_\tau$ at the detector is related
to the $\nu_\mu$ flux at the source through,
\begin{eqnarray}
     \phi_{\nu_\mu} &=& \left(1-\frac{1}{2}\sin^22\theta\right)
     \phi_{\nu_{\mu}}^0,\\
     \phi_{\nu_\tau} &=& \frac{1}{2}\sin^22\theta \
     \phi_{\nu_{\mu}}^0,
\end{eqnarray}
and $\theta$ will in general differ from $\theta_v$ if
$\meff\neq0$.

\subsection{Proposed Measurement}

Here we investigate the possibility of measuring $\theta$ using
current or future ultra-high-energy--neutrino experiments such
as Auger \cite{Auger2009} and ANITA \cite{ANITA}.
It is in principle possible to discriminate $\nu_\tau$
from $\nu_\mu$ by separately measuring the Earth-skimming events ($\nu_\tau$) and almost horizontal events originating 
in air ($\nu_\mu$).

We assume that a given experiment detects $N_\nu^{\rm tot}$
neutrino events.  This quantity is for the {\it total} neutrino
and antineutrino flux; i.e., $N_\nu^{\rm tot} = N_{\nu_\mu} +
N_{\nu_\tau}$ (here $\nu$ represents both neutrinos and
antineutrinos).  If the flavor democracy
expected from vacuum mixing is realized, then one expects
$N_{\nu_\mu} =  N_{\nu_\tau} = N_{\nu}^{\rm tot} / 2$.
The number of neutrino events at the detector is related to the
flux through,
\begin{equation}
     N_{\nu} = \int_{E_{\rm min}}^{E_{\rm max}} dE\, \phi_\nu (E)
     {\Xi}(E),
\end{equation}
where $\Xi(E)$ is the detector exposure to neutrinos in units of
cm$^2$~s~sr, and it generally depends on neutrino energy.
Here we assume $\phi_\nu^{\rm tot} = \phi_{\nu_\mu}^0 = K E^{-2}$ with
a normalization constant $K$, $E_{\rm min} = 2 \times 10^{17}$ eV, and
$E_{\rm max} = 2\times 10^{19}$ eV.
This provides a good approximation for the spectrum of cosmogenic
neutrinos (e.g., Ref.~\cite{Yuksel2007}).
For simplicity we further assume that the detector exposure is independent of energy; the Auger exposure indeed depends on neutrino energy
only weakly~\cite{Auger2009}.
Therefore, the total number of neutrino events is given by
\begin{equation}
     N_\nu^{\rm tot} = \frac{K {\Xi}}{E_{\rm min}},
\label{eq:N total}
\end{equation}
and the number of $\nu_\tau$ events is given by
\begin{eqnarray}
     N_{\nu_\tau} &=& \int_{E_{\rm min}}^{E_{\rm max}} dE\, \frac{1}{2}\sin^2
     2\theta \phi_{\nu_\mu}^0(E) \Xi
\nonumber\\ &=&
     \frac{K \Xi}{2}\int_{E_{\rm min}}^{E_{\rm max}} dE\, E^{-2} \sin^2
     2\theta,
\nonumber\\&=&
     \frac{N_\nu^{\rm tot} E_{\rm min}}{2} \int_{E_{\rm min}}^{E_{\rm max}}
     dE \, E^{-2} \sin^2 2\theta,
\label{eq:N nutau}
\end{eqnarray}
where we used Eq.~(\ref{eq:N total}) in the last equality.

\begin{figure}
\includegraphics[width=8.6cm]{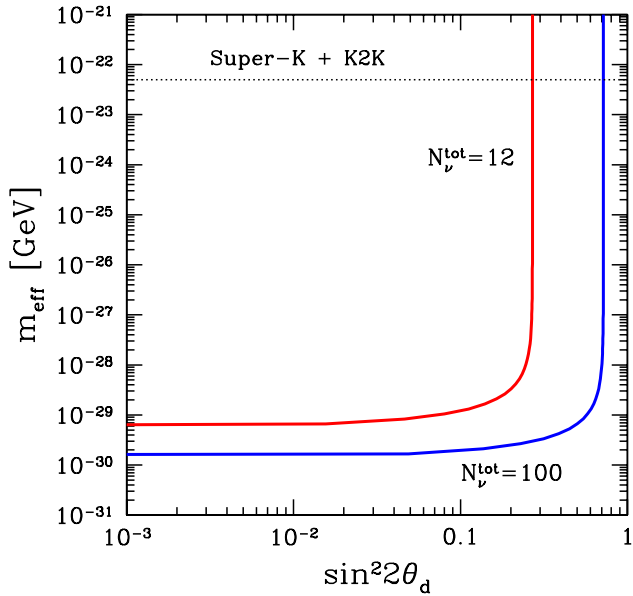}
\caption{Sensitivity on ($m_{\rm eff}$, $\sin^22\theta_d$) plane of
  future experiments that would yield $N_\nu^{\rm tot} = 12$ and 100
  total neutrino events.}
\label{fig:GZK}
\end{figure}

To investigate the sensitivity of a given experiment, we assume a
null detection of new physics; i.e., the result of $N_{\nu_\tau}$
is consistent with the standard expectation $N_{\nu}^{\rm tot} / 2$
within statistical errors (we do not take systematic uncertainties
into account).
This will reject a certain range of parameter space for ($m_{\rm eff}$,
$\sin^2 2\theta_d$).
More specifically, to obtain 95\% C.L. ($2\sigma$) limits for
these parameters, we solve
\begin{equation}
     N_{\nu_\tau} > \frac{N_\nu^{\rm tot}}{2} - 2 \sqrt{\frac{N_\nu^{\rm
     tot}}{2}},
\end{equation}
for $m_{\rm eff}$ and $\theta$, using Eq.~(\ref{eq:N nutau}) for the
left-hand side.
In Fig.~\ref{fig:GZK}, we show the sensitivity of detectors that
are expected to collect 12 and 100 neutrino events\footnote{The
  current Auger exposure is $\Xi \sim
  10^{16}$~cm$^2$~s~sr~\cite{Auger2009}, and an optimistic
  estimate for the flux of cosmogenic neutrinos is
  close to the Waxman-Bahcall bound~\protect\cite{WB}, $E^2 \phi_\nu(E)
  \sim 10^{-8}$ GeV
  cm$^{-2}$ s$^{-1}$ sr$^{-1}$~\protect\cite{Yuksel2007}. Therefore, from
  Eq.~(\ref{eq:N total}), we expect $N_\nu^{\rm tot} \alt 1$, which is
still consistent with nondetection by Auger.} (total) and
that also have a $\nu_\tau$-identification capability.
If the true values of $m_{\rm eff}$ and $\sin^22\theta_d$ are above
these curves, then we will see an anomalously suppressed
$\nu_\tau$ flux compared with the standard mixing scenario.
We also show the current upper limit on $m_{\rm eff}$ obtained from
the combined analysis of Super-K and K2K data performed in
Ref.~\cite{Gonzalez-Garcia2004}.
One can see from this Figure that by detecting cosmogenic
neutrinos and by studying their flavor content, one can largely
improve the current sensitivity to $m_{\rm eff}$ and $\theta_d$,
quantifying further the suggestion of Ref.~\cite{Hooper:2005}.
We also note that a weaker sensitivity, albeit still much better
than the current sensitivity, may be achieved with neutrinos of
slightly lower energies \cite{MCGG}.

\section{Matter effects in atmospheric neutrino oscillations in the presence of a dark-energy coupling}
\label{resonance}

We now turn our attention to Lorentz/{\it CPT}-violating effects
in electron-neutrino oscillations, showing here
that novel effects may arise with DE-neutrino coupling as
neutrinos propagate through the Earth.  These effects may
allow us to access with atmospheric neutrinos regions of the
DE-neutrino--coupling parameter space significantly below those
currently probed.  In this section, we consider two-flavor and
three-flavor oscillations.

As neutrinos travel through matter, there is an additional
contribution to oscillations from the matter potential $\sqrt{2}
G_F N_e$ (where $G_F$ and $N_e$ are, respectively, the Fermi
constant and electron density) relevant
if electron neutrinos are involved.  Recalling that the matter
potential is $\gtrsim 10^{-22}$~GeV, the vacuum-mixing term
$\Delta m^2/2E$ is small for neutrino energies $\gtrsim10$~GeV.
The mixing matrix Eq.~(\ref{eqn:DEnumixinggeneral}) then becomes
for $\nu_e$-$\nu_\mu$ mixing (neglecting the overall factor of $1/2$,
the directional dependence, and the phase $\eta$),
\begin{equation}
     \left( \begin{matrix}  -\meff \cos 2 \theta_d  +\sqrt{2}
     G_F N_e & \meff \sin 2\theta_d \\ \meff \sin 2 \theta_d &
     \meff \cos 2 \theta_d  -\sqrt{2} G_F N_e \end{matrix}
     \right).
\label{eqn:mattermatrix}
\end{equation}
Note that here, both the DE term and the matter potential change sign
for antineutrinos, unlike the usual MSW effect, in which the
vacuum term does not change sign. Unlike MSW mixing, there
is essentially no energy dependence, at sufficiently high
energies, in this mixing matrix.

To see when DE-induced mixing may be significant,
recall that the value of the matter potential is
$\sqrt{2} G_F N_e = 7.6 \times10^{-14}\, Y_e
(\rho/\mathrm{g}~\mathrm{cm}^{-3})~\mathrm{eV}$.  The Earth core
has average density $\rho_{\mathrm{core}} = 11.83\,
\mathrm{g}~\mathrm{cm}^{-3}$ and electron fraction $Y_e^{\mathrm{core}} =
0.466$, while the mantle has average density $\rho_{\mathrm{mantle}} =
4.66\, \mathrm{g}~\mathrm{cm}^{-3}$ and $Y_e^{\mathrm{mantle}} = 0.494$,
with the surface layer of the Earth having density as low as
2.6~$\mathrm{g}~\mathrm{cm}^{-3}$. The matter potential is thus
about $10^{-13}$ eV, so the effects of DE-induced mixing may be
manifest for $\meff$ around $10^{-22}$ GeV, well below current
upper limits. In the absence of matter, as discussed in
Ref.~\cite{bargeretal}, it is possible to obtain a resonance when all
mixing angles involved are maximal
\begin{equation}
\frac{\Delta m^2}{2E} \cos2\theta_v+\meff \cos2\theta_d=0 .
\end{equation} 
Here, in the presence of matter and at high energies, a resonance can
occur for a small mixing angle $\theta_d$ when
\begin{equation}
\meff \cos2\theta_d = {\sqrt{2}} G_F N_e .
\end{equation} 
The presence of a resonance is thus entirely determined by the densities
encountered along the path and the DE coupling parameters, with no (or
very weak) energy dependence at high energies.

\begin{figure}
\includegraphics[width=8.0cm,clip=true]{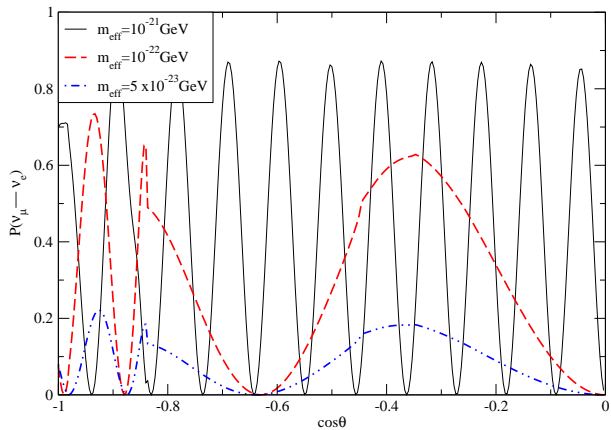}
\caption{Oscillation probability as a function of zenith angle for
 atmospheric neutrinos of $E=50$ GeV, obtained with $\theta_d = \pi/4$.}
\label{fig:meff}
\end{figure}

To illustrate the possibilities, we integrate the
neutrino-propagation equation (including the small vacuum-mixing term)
to calculate the $\nu_\mu$-to-$\nu_e$ transition probability as
a function of (cosine of) the zenith angle for atmospheric
neutrinos propagating through the Earth.  We use the density profile
of the Earth as given by the PREM model \cite{Dziewonski:1981xy}.
Figure~\ref{fig:meff} shows the results for two-flavor
oscillations for different values of
$\meff$ for $\theta_d=\pi/4$.  When $\meff \gg \sqrt{2} G_F
N_e$, the oscillation probability is determined almost entirely
by the DE term;  there are regular large-amplitude variations of the
oscillation probability as a function of zenith angle.  As
$\meff$ decreases to values comparable to $\sqrt{2} G_F N_e$,
the oscillation probability decreases, and the oscillation
length is seen to differ for trajectories that do ($\cos\theta
\lesssim -0.8$) and do not ($\cos\theta \gtrsim -0.8$) pass
through the core.

\begin{figure}[t]
\includegraphics[width=8.cm,clip=true]{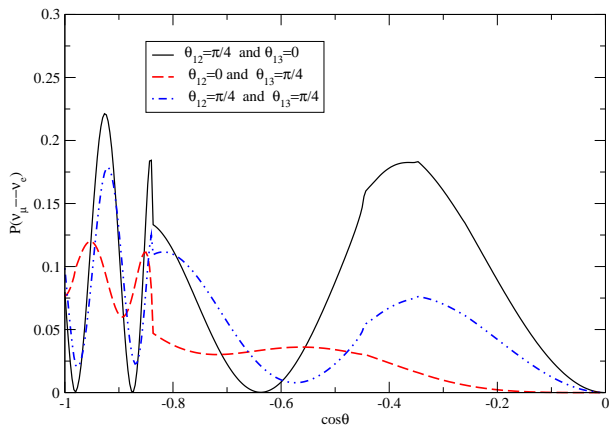}
\caption{Oscillation probability as a function of zenith angle for
 atmospheric neutrinos of $E=50$ GeV, obtained for three-flavor
 oscillations with various values of $\theta_{12}^d$ and $\theta_{13}^d$
 and $m_{\rm eff} = 5 \times 10^{-23}$ GeV.}
\label{fig:2angles}
\end{figure}

In Fig.~\ref{fig:2angles} we show the oscillation
probabilities for $\meff=5\times 10^{-23}$ GeV for three-flavor mixing
in the Lorentz-violating sector. The case where
$\theta_{13}^d=\pi/4$ and $\theta_{12}^d=0$ also corresponds to an
effective two-flavor scenario, just like the previous
results. It leads, however, to a very different behavior due to
the different contribution of the standard neutrino
oscillations. The case where $\theta_{13}^d=\theta_{12}^d=\pi/4$
corresponds to a full three-flavor oscillation scenario.
We have also studied the effects for other values of the mixing
angles and the same features remain present.  A non-zero value
of $\theta_{13}^v$ for standard neutrino oscillations leads
to similar features in the zenith-angle distribution.  However,
the effects are extremely small at the high energies considered
here, orders of magnitude below those coming from the
Lorentz-invariance--violating terms.

\section{Directional dependence}
\label{sec:direction}

While detection of {\it CPT}/Lorentz-violating effects would be
spectacular---it would imply new physics regardless of whether
it is DE-related or not---the real ``smoking gun'' for a DE
effect would be the directional dependence, $\propto (1-\vecv
\cdot \hat{\vecp})$, of neutrino-oscillation parameters.  Given
that our peculiar velocity with respect to the CMB rest frame is
$10^{-3}$ times the speed of light, the magnitude of this effect
is going to be suppressed relative to the other effects,
discussed above, of a DE-neutrino interaction.  Statistics
well beyond the reach of current and forthcoming neutrino
experiments will be required to detect this effect.  Still, it
is worth keeping in mind for future generations of experiments.

It may also be worth searching for such a directional
dependence in current data, just in case there is a DE-neutrino
coupling that is manifest in ways different than we have
foreseen here.  For example, if DE somehow produces Lorentz
violation through a modification of the kinetic term in the
Dirac equation, the energy dependence of the mixing induced by
Lorentz/{\it CPT}-violation could be different
\cite{Kostelecky2004}.  We therefore work out in this Section
expressions for the factor $\vecv \cdot \hat{\vecp}$ to aid
experimentalists who may wish to look for direction-dependent
effects in their neutrino (or other) data.

To proceed, we first set our coordinate system.
We set the origin at the center of Earth and align the $z$ axis along
the rotational axis of Earth, so that the north pole has positive $z$
coordinate.
We set the $x$ axis along the direction to the Sun at vernal equinox.
Since the Sun moves eastbound, its position at summer solstice aligns
with the $y$-axis.
We can thus represent the seasonal shift by an azimuthal angle $\varphi$,
where $\varphi = 0$, $\pi/2$, $\pi$, and $3\pi/4$ for vernal equinox,
summer solstice, autumn equinox, and winter solstice, respectively.
Note also that the orbital plane of the Sun is inclined from the
$x$-$y$ plane by $\theta_{\rm inc} = 23.5^\circ$~\cite{Binney1998}.

The Sun is moving with respect to the CMB rest frame with a speed
of $v_{\odot} = 369$ km s$^{-1}$ towards the direction $\alpha =
168^\circ$, $\delta = -7.22^\circ$~\cite{Lineweaver1996}, where $\alpha$
is right ascension and $\delta$ is declination of the celestial
coordinates~\cite{Binney1998}.
In our coordinates, the velocity of the Sun is $\vecv_\odot = v_\odot
(\cos\delta \cos\alpha, \cos\delta \sin\alpha, \sin\delta) = (-358,
76.1, -46.4) ~ \mathrm{km~ s^{-1}}$.  The Earth is moving around
the Sun with average orbital speed of $V = 29.8$ km s$^{-1}$.
Thus, the velocity of the Earth with respect to the CMB rest
frame is
\begin{eqnarray}
 \vecv_{\oplus} &=& \vecv_\odot +
  V\left(
  \begin{array}{c}
   \sin\varphi\\
   -\cos\varphi \cos\theta_{\rm inc}\\
   -\cos\varphi \sin\theta_{\rm inc}\\
  \end{array}
  \right)
  \nonumber\\
 &=& \left(
     \begin{array}{dc}
     -358&+29.8\sin\varphi\\
      76.1&-27.3\cos\varphi\\
      -46.4&-11.9\cos\varphi\\
     \end{array}
     \right)~\mathrm{km~s^{-1}}.
 \label{eq:Earth velocity}
\end{eqnarray}
We neglect the contribution from the rotation of Earth
($\lesssim0.5$~km~sec$^{-1}$) to our velocity with respect to the CMB
rest frame.

Now we evaluate the direction of the neutrino beam $\hat{\vecp}$.
We suppose that the beam runs from some point A on the Earth's
surface to another point B on its surface (or {\it vice versa}).
It should be straightforward to generalize the arguments below so
that extraterrestrial neutrino production can be taken into account.
We set the origin of time coordinate $T$ to ``noon'' (i.e., when
the Sun reaches highest) at the point A.  Therefore, the
positions of A and B in our coordinate are
\begin{eqnarray}
\vecx_{\rm A} &=& R_\oplus
 \left(
 \begin{array}{c}
  \cos\phi_{\rm A}\cos(\omega T_{\rm A}+\varphi) \\
  \cos\phi_{\rm A}\sin(\omega T_{\rm A}+\varphi) \\
  \sin\phi_{\rm A}
 \end{array}
 \right),
 \label{eq:x_A}
\\
\vecx_{\rm B} &=& R_\oplus
 \left(
 \begin{array}{c}
  \cos\phi_{\rm B}\cos(\omega T_{\rm A}+\Delta\lambda+\varphi) \\
  \cos\phi_{\rm B}\sin(\omega T_{\rm A}+\Delta\lambda+\varphi) \\
  \sin\phi_{\rm B}
 \end{array}
 \right),
 \label{eq:x_B}
\end{eqnarray}
where $R_\oplus$ is the radius of the Earth, $\omega$ is the rotational
frequency ($2\pi / {\rm day}$), $T_{\rm A}$ is the time at the position A
relative to noon, $\phi_{\rm A,B}$ is the geometric latitude of the
points A and B, and $\Delta \lambda = \lambda_{\rm A} - \lambda_{\rm B}$
is the difference of the geometric longitude.
The quantity $\Delta\lambda$ appears in $\vecx_{\rm B}$ because we
measure the time (for both A and B) with respect to noon of the point A,
so the time difference is given by the latitude difference (note also
that the longitude increases to the west).
The direction of the neutrino beam $\hat{\vecp}$ is then proportional
to $\vecx_{\rm B} - \vecx_{\rm A}$ with proper normalization as
\begin{widetext}
\begin{equation}
 \hat{\vecp}= \frac{1}{\sqrt{2(1-\cos\phi_{\rm A}\cos\phi_{\rm
  B}\cos\Delta\lambda - \sin\phi_{\rm A} \sin\phi_{\rm B})}}
  \left(
  \begin{array}{c}
   \cos\phi_{\rm B}\cos(\omega T_{\rm A}+\Delta\lambda+\varphi) -
    \cos\phi_{\rm A} \cos(\omega T_{\rm A}+\varphi) \\
   \cos\phi_{\rm B}\sin(\omega T_{\rm A}+\Delta\lambda+\varphi) -
    \cos\phi_{\rm A} \sin(\omega T_{\rm A}+\varphi) \\
   \sin\phi_{\rm B} - \sin\phi_{\rm A}
  \end{array}\right).
  \label{eq:p hat}
\end{equation}
\end{widetext}

Therefore, by combining Eqs.~(\ref{eq:Earth velocity}) and (\ref{eq:p
hat}), we obtain the directional factor $1 - \vecv_\oplus \cdot
\hat{\vecp}$.  Since it is a scalar quantity, the final result
does not depend on the choice of the coordinate system.

\section{Theoretical Implications}
\label{implications}

Before closing, we discuss, for illustration, the implications
of a measurement of a particular value of the
Lorentz-invariance--violating effective mass parameter $\meff$
in terms of a specific model of DE-neutrino coupling.  

Perhaps the simplest interaction of this kind has the form,
\begin{equation}
 \mathcal L_{\rm int} = - \lambda_{\alpha \beta} \frac{\partial_\mu
 \phi}{M_\ast} \bar\nu_\alpha \gamma^{\mu} (1-\gamma_5) \nu_\beta ,
\end{equation}
where $\phi$ is a quintessence field, $\lambda_{\alpha \beta}$
is a coupling-constant matrix, and $M_\ast$ is some mass scale.
Thus, $a_L^\mu \sim \lambda \dot \phi(t) l^\mu / M_\ast$, and $m_{\rm
eff} \sim \Delta\lambda \dot \phi(t) / M_\ast$, where $\Delta \lambda$
is the difference between eigenvalues of the $\lambda$ matrix.
For quintessence, one expects $\dot \phi
\sim M_{\rm Pl} H_0 (1+w)^{1/2}$ (e.g., Ref.~\cite{Caldwell:2009ix}),
where $M_{\rm Pl}$ is the Planck energy scale.  In this case,
the mass scale $M_\ast$ corresponding to a given $\meff$ is
\begin{equation}
     M_\ast \simeq 10^{6} \, (\Delta \lambda)
     \left(\frac{1+w}{0.01}\right)^{1/2}
     \left(\frac{\meff}{10^{-30}\,\mathrm{GeV}} \right) ~\mathrm{GeV}, 
\end{equation}
to the mass scale that controls the DE-neutrino interaction.
The ultra-high-energy $\nu_\mu$-$\nu_\tau$ oscillation effects
we have discussed thus probe up to mass scales
$M_\ast\sim10^6$~GeV.  The $\nu_e$-$\nu_\mu$ oscillations
induced by the matter effects we discussed probe up to mass
scales $M_\ast\sim100$~MeV.  

\section{Conclusions}
\label{sec:conclusion}

We studied the implications of an interaction between dark
energy and neutrinos for neutrino oscillations.  The most
general Lorentz/{\it CPT}-violating term induced by dark energy (DE) takes
the form $(a_L)^\mu \bar\nu \gamma_\mu (1-\gamma_5) \nu$, where
$(a_L)^\mu$ is a four-vector normal to the CMB rest frame.  This
introduces a new source for neutrino oscillations that are energy
independent and different for neutrinos and antineutrinos.  Furthermore,
the motion of the Earth with respect to the cosmic rest frame induces a
directional dependence in the oscillation probabilities.

The current best limits to the DE-neutrino coupling we
considered are obtained from atmospheric- and
accelerator-neutrino experiments for $\nu_\mu$-$\nu_\tau$
mixing, and from solar and reactor experiments for
$\nu_e$-$\nu_\mu$ mixing.  However, the higher the neutrino
energy, the more prominent the effect of the DE-neutrino
interaction.  We therefore considered 
in this paper cosmogenic ultra-high-energy (energies of
$10^{17}$--$10^{19}$ eV) neutrinos produced by the interaction
of ultra-high-energy cosmic rays with CMB photons.
We showed that future experiments targeting these neutrinos
will improve the sensitivity to a DE-neutrino interaction by seven
orders of magnitude, down to $m_{\rm eff} \sim 10^{-30}$ GeV
compared with the current upper bound $m_{\rm eff} \alt 5\times
10^{-23}$ GeV (Fig.~\ref{fig:GZK}).  This
corresponds to a sensitivity to an energy scale as large as
$\sim$10$^6$ GeV for the DE-neutrino interaction.  We then
showed that the interplay of DE- and
matter-induced neutrino mixing could induce a novel zenith-angle
dependence for $\nu_e$ oscillations in atmospheric
neutrinos.  This effect may extend the sensitivity to Lorentz/{\it
CPT}-violating parameters in the $\nu_e$ by roughly three orders
of magnitude.

The real smoking gun of a DE-neutrino
interaction (as opposed to some other origin for Lorentz/{\it
CPT} violation) would be a directional dependence of the oscillation
probabilities.  The notion that Lorentz violation
may give rise to a directional dependence is not new (e.g.,
Ref.~\cite{Kostelecky:2004hg}) and searches for directional
dependence in neutrino experiments have already been carried out
(e.g., Ref.~\cite{Adamson:2008ij}), but prior work has considered
Lorentz-violating parameters introduced in an {\it ad hoc}
manner and/or tested for direction-dependent effects in a
Sun-centered inertial frame.  We emphasize here that cosmic
acceleration suggests that we seek a specific form of Lorentz
violation, that where the preferred frame is aligned
with the CMB rest frame.
Even though such a signal is expected to be small, it is still
worth seeking in existing and future experimental data.

We have not discussed specific models for a
DE-neutrino interaction, beyond an illustrative toy
model, but it may be interesting to do so (see also
Ref.~\cite{stephon}).  The
theoretical motivation to expect such a coupling may admittedly
be slim.  However, we are at square one in our understanding of
DE, and such a coupling is no less likely to be
expected, perhaps, than any of the many other
manifestations of new cosmic-acceleration physics that
have been considered.  Discovery of Lorentz/{\it CPT}-violating
effects would be extremely important, even if not attributable
directly to dark energy.  A directional dependence, if
discovered, would be absolutely remarkable, as it would provide
moreover clear evidence that there is more to cosmic
acceleration than simply a cosmological constant.

\acknowledgments

We thank Alexander Friedland and Stephon Alexander for useful
discussions, and we acknowledge the hospitality of the Aspen
Center for Physics.  This work was supported by the Sherman
Fairchild Foundation (SA), DoE DE-FG03-92-ER40701 (MK), and NSF
grant PHY-0555368 (IM).

\end{document}